\documentclass[amsmath,amssymb,aps,pra,twocolumn,superscriptaddress]{revtex4-1}
\usepackage{graphicx}% Include figure files
\usepackage{dcolumn}% Align table columns on decimal point
\usepackage{bm}% bold math
\usepackage{hyperref}% add hypertext capabilities
\usepackage[mathlines]{lineno}% Enable numbering of text and display math
\usepackage{lineno}
\usepackage{graphicx}% Include figure fileshttps://www.overleaf.com/project/5fda550c473e7229bc0f2258
\usepackage{amsfonts} %matrices
\usepackage{dsfont}
\usepackage{braket} %bras, kets, etc
\usepackage{hyperref} %hyperlinks y referencias
\usepackage{xcolor}
\usepackage{amsmath}

\begin{document}

\preprint{APS/123-QED}

\title{Role of mechanical effects on the excitation spectra of microwave-dressed Rydberg states in a  cold atomic cloud.}

\author{H. Failache}
\email{heraclio@fing.edu.uy}
\author{J.A. Muniz}
\author{L. Velazco}
\author{D. Talento}
\author{A. Lezama}

\affiliation{Instituto de F\'{\i}sica, Facultad de Ingenier\'{\i}a,
Universidad de la Rep\'{u}blica,\\ J. Herrera y Reissig 565, 11300
Montevideo, Uruguay}

\date{\today}% It is always \today, today,
             %  but any date may be explicitly specified

\begin{abstract}
We explore the excitation spectra of cold $^{87}$Rb atoms to the 55D$_{3/2}$ Rydberg state in the presence of microwave (MW) radiation as a function of MW frequency. The spectra reveal several features around the transition-frequencies between adjacent Rydberg states. We argue that some of these features are indicative of variations in the Rydberg excitation probability while others result from the removal of atoms from the cold cloud as a consequence of a MW induced strong dipole-dipole inter-atomic force. Our claim is supported by experimental observations and theoretical modeling.
\end{abstract}

%\keywords{Suggested keywords}%Use showkeys class option if keyword
                              %display desired
\maketitle

%\tableofcontents

\section{Introduction}

Rydberg atoms - in which an electron has been promoted to a high energy state close to the ionization limit - have long attracted the interest of the atomic physics community \citep{Gallagher88}. They possess universal characteristics largely independent of the specific atomic element. Their wave-functions and energy levels can be calculated with extreme precision thanks to the quantum defect theory \citep{Li03} and this, in turn, allows the precise evaluation of most atomic properties \citep{Shao24}. Arguably the most relevant property of Rydberg atoms are the large values of the electric-dipole matrix elements between neighboring electronic states resulting in a very large atomic polarizability. As a consequence, Rydberg atoms are extremely sensitive to electromagnetic (EM) fields either externally imposed or due to the presence of other Rydberg atoms.\\

The extreme sensitivity of Rydberg atoms to EM fields has long been used as a sensitive tool for the detection and measurement of ambient fields \citep{Sedlacek12}. It has also allowed the realization of fundamental tests of quantum theory in the nearly ideal configuration of a single two-level atom in the presence of a single photon in a unique EM mode \citep{Gleyzes07}.\\ 

In addition, the sensitivity of Rydberg atoms to EM fields results in the possibility of strong dipole-dipole interaction (DDI) between Rydberg atoms at large distances (several micrometers). Such interaction can be exploited as a means to couple two otherwise independent atomic systems. It also results in a significant shift of atom-pair energy levels with respect to infinitely separated atoms. As a consequence, two sufficiently close atoms cannot be both excited to the same Rydberg state via narrowband excitation, an effect known as the Rydberg blockade \citep{Low12,Urban09,Wu23}. Both effects, distant atoms coupling and Rydberg blockade, have attracted considerable interest in recent years as useful tools for quantum information processing \citep{Saffman10,Lim13,Adams19,Zhang20}.\\

In cold atom samples the inter-atomic forces resulting from the DDI can be strong enough to accelerate the atoms to speeds beyond the thermal velocity and possibly expel the atoms from the trapping volume \citep{Li05,Amthor07,Amthor07PRL,Faoro16,Park16,Thaicharoen16}. Also, many-body interactions may result in modifications of the spatial atomic distribution \citep{Amthor07,Schauss12}.\\ 

In most experiments concerning Rydberg states, atom ionization is employed for detection. It is a very sensitive technique which allows single atom detection and Rydberg energy level identification. However, it requires the presence of electrodes near the observed volume and a Channeltron detector inside the vacuum chamber.\\ 

An alternate all-optical Rydberg detection technique is provided by electromagnetically induced transparency (EIT) \citep{Fleischhauer05}. It relies on two-photon excitation of the Rydberg transition using as an intermediate state  a lower atomic level. When the two-photon resonance condition is met, the absorption of either exciting beams is reduced. EIT has the advantage that the necessary setup can be entirely placed outside the vacuum chamber. It also allows a good spatial selectivity since the probing lasers can be tightly focused on the sample. It is however less sensitive than ionization since it requires a significant optical density of the atomic sample. EIT has been extensively used in cold Rydberg atoms experiments \citep{Weatherill08,Tanasittikosol11,Mack11}.\\ 

Recent experiments exploit, as a spectroscopic tool, the atom-loss resulting from Rydberg excitation in cold atomic clouds. The detection is performed by monitoring the number of atoms remaining in the cloud after Rydberg excitation. This can be achieved by atomic cloud absorption measurement \citep{deHond20} or by monitoring the cloud brightness  in a magneto optical trap (MOT) \citep{Cao22,Halter23,Duverger24,Kondo24}. \\

Electric-dipole-allowed atomic transitions between Rydberg states often occur in the microwave (MW) frequency range \citep{Dyubko95}. In consequence, MW radiation was extensively used for probing and interacting with Rydberg atoms \citep{Park16}.\\

In this article we explore the interaction of MW radiation with Rydberg atoms excited and detected via EIT. The configuration corresponds to a three photon (two optical and one MW) excitation scheme. Alternatively, our experiment can be described as the EIT spectroscopy of Rydberg atoms dressed by the MW radiation. We have recorded the Rydberg excitation spectra as a function of MW frequency and observed a variety of features some occurring at transition frequencies between adjacent Rydberg states and others for MW frequencies detuned from atomic  transitions. The latter are assigned to Rydberg atom pairs whose energy levels are modified by DDI. Interestingly, some of the features observed as a function of the MW frequency correspond to an increased transparency of the sample for the IR probing beam. We argue that such increased EIT cannot be just the consequence of variations of the Rydberg excitation probability of atom pairs but can be understood as resulting from the strong mechanical effect arising from the DDI within the pair. Our argument is supported by experimental evidence and consistent with a dressed Rydberg atom-pair modeling. \\       

\section{Experiments} \label{experimentos}
\subsection{Experimental setup} \label{setup}

The atomic sample consists of $^{87}$Rb  atoms cooled and trapped in a MOT. The temperature of the cloud is $\sim$100 $\mu$K and its density $\rho$ = 10$^{8}$ atoms/mm$^{-3}$, corresponding to an average inter-atomic distance $\rho^{-1/3} \sim$ 2 $\mu$m. The atomic cloud has a typical diameter of $\sim$  300 $\mu$m. We use a standard MOT with cooling beams tuned near the 5S$_{1/2}$ (F=2) $\rightarrow$ 5P$_{3/2}$ (F=3) transition and a repumping beam tuned to the 5S$_{1/2}$ (F=1) $\rightarrow$ 5P$_{1/2}$ (F=1) transition.\\

\begin{figure} [htbp]
  \centering
  \includegraphics[width=1\linewidth]{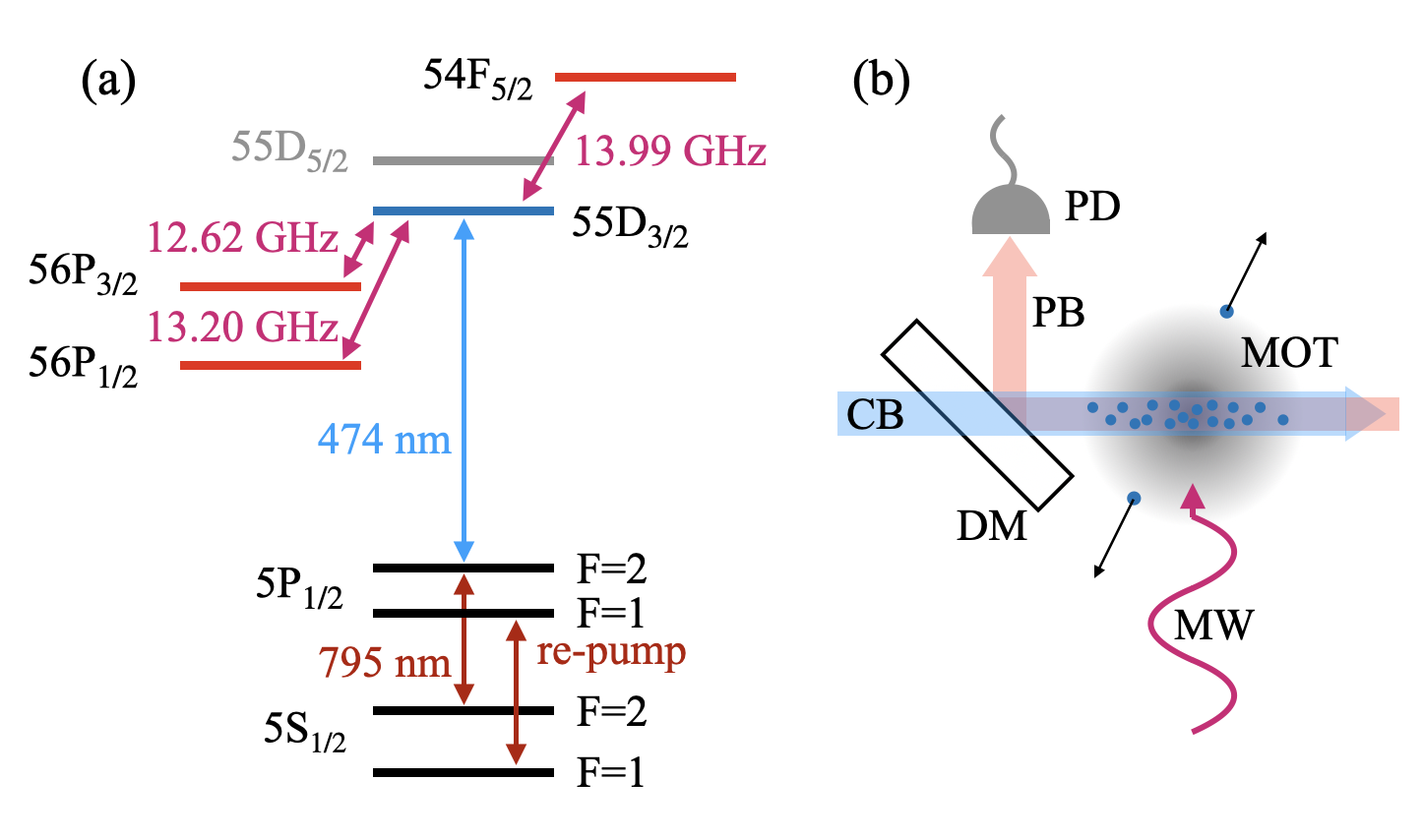}
\caption{a) \label{esquemas} Relevant energy level scheme. b) Spectroscopic setup [MOT: cold atom cloud, DM: dichroic mirror, PD: photodetector, MW: microwave radiation, CB: coupling beam (blue), PB: probe beam (IR)].}
\end{figure}

Some cold atoms are excited to the Rydberg 55D$_{3/2}$ states using the two-photon ladder excitation process illustrated in Fig. \ref{esquemas}.a. An infrared (IR) probe laser ($\lambda \simeq$ 795 nm) resonantly excites ground-state atoms to the 5P$_{1/2}$ (F=2) state. A blue coupling laser ($\lambda \simeq$ 474 nm) promotes the atoms to the 55D$_{3/2}$ Rydberg state. Both radiations originate from extended cavity diode lasers. The IR laser was stabilized to the transition [5S$_{1/2}(F=2) \rightarrow\ $5P$_{1/2}(F=2)$] by saturated absorption spectroscopy. The blue laser is stabilized to the two-photon transition from the ground state to the Rydberg state 55D$_{3/2}$ in a hot Rb reference vapor cell, using the IR laser as a first excitation step. The EIT of the IR light provides the error signal for stabilization.  At the cold cloud position both beams have the same vertical linear polarizations. The two beams counter-propagate through the cold atomic cloud with a beam waist of approximately 50 $\mu$m. The power of the IR and blue beams were 50 nW and 5 mW respectively, which correspond to a Rabi frequency of 7 MHz for the 5S$_{1/2}(F=2) \rightarrow\ $5P$_{1/2}(F=2)$ transition and 13 MHz for the 5P$_{1/2}$(F=2) $\rightarrow$ 55D$_{3/2}$ transition.

A double-ridged horn antenna, located outside the stainless steel vacuum chamber, generates vertically polarized MW radiation. The typical MW electric field amplitude at the position of the MOT was estimated to be ~0.4 V/m, which corresponds to a Rabi frequency of 10 MHz for the 55D$_{3/2}$ $\rightarrow$ 54F$_{5/2}$ microwave transition. The MW frequency was tuned over a $\sim$2 GHz interval including the transition frequencies between the state 55D$_{3/2}$ and the nearby 54F$_{5/2}$, 56P$_{3/2}$ and 56P$_{1/2}$ states (see Fig.\ref{esquemas}.a).\\

\subsection{Results} \label{measurements}

Before investigating the effect of the MW field, we verify the excitation of Rydberg atoms by scanning the IR probe laser and monitoring its transmission through the atomic cloud for a fixed frequency of the blue laser. A visible increase in IR transmission was observed when the two-photon resonance condition was met between the ground and the 55D$_{3/2}$ Rydberg state ($\sim$40 \% transmission increase with FWHM $\approx 20$ MHz ) . We loosely designate this transparency increase as EIT. As a coherent phenomenon, EIT requires a high degree of mutual coherence between the involved lasers which is not be met in our setup. Nevertheless, the promotion of atoms to the Rydberg state implies, in the steady state, a reduction of the ground-state population which is the main parameter controlling the IR beam absorption.\\ 

\begin{figure} [htbp]
  \centering
\includegraphics[width=1\linewidth]{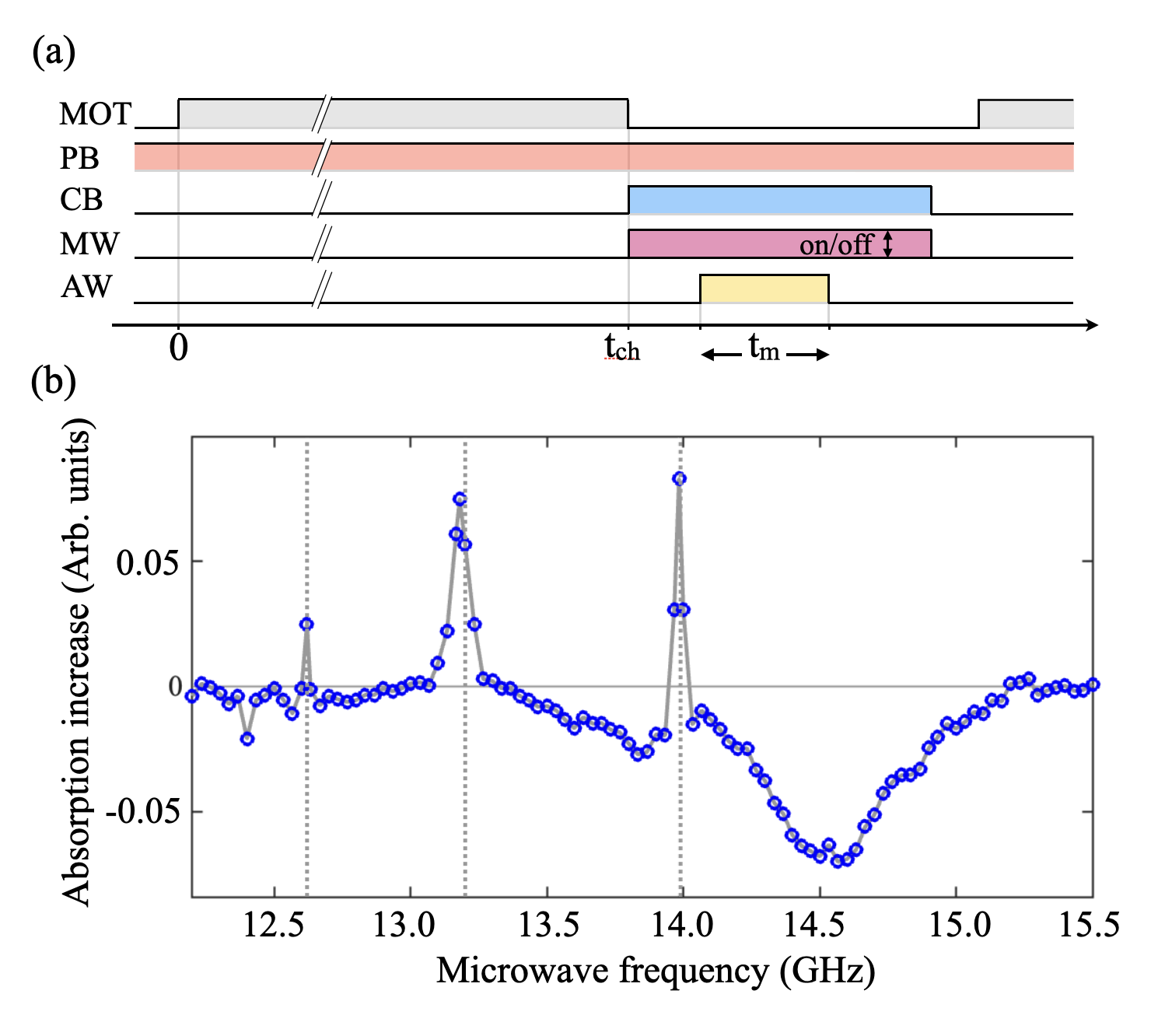}
\caption{\label{absorcion} a) Time sequence (MOT: cooling beams, PB: probe beam, CB: coupling beam, MW: MW field, AW: absorption averaging time-window). b) Probe beam EIT variation as a function of MW frequency. The dotted lines indicate the theoretical frequencies corresponding to  transitions to the nearest Rydberg levels shown in Fig. \ref{esquemas} \citep{Sibalic17}.}
\end{figure}

To investigate the effect of the MW field, we have measured the change in probe beam absorption as a function of the MW frequency while keeping the probe and coupling optical field frequencies locked to the two-photon resonance condition. Figure \ref{absorcion}.a presents the corresponding time sequence. The IR probe beam was always present. The cold atoms were captured in the MOT during a charging time $t_{ch}$ = 1 s after which the cooling beams were turned off (releasing the cold atomic cloud). Simultaneously, the blue light was turned on. The transmission of the IR probe was recorded 500 $\mu$s after turning on the blue light (allowing the EIT signal to approach its maximum) and averaged over an interval $t_{m}$ = 1 ms.
The IR transmission was recorded for alternate realizations of the cold cloud with and without the MW field present in simultaneity with the coupling beam. The signal shown in Fig. \ref{absorcion}.b is the difference between these two conditions. Each point corresponds to the averaging of 100 realizations of the MOT. The base line in \ref{absorcion}.b represents the EIT signal in the absence of MW.\\

As expected, the spectrum shows narrow absorption peaks at MW frequencies corresponding to the transition to the nearby energy levels indicated in Fig. \ref{esquemas}.a. However, the most striking feature of the spectrum in Fig.\ref{absorcion}.b is arguably the increase in transparency beyond the level reached by optical-only Rydberg excitation (zero of the vertical axis) which is observed over several hundreds of MHz to the blue of the 55D$_{3/2} \rightarrow$ 54F$_{5/2}$ transition frequency.\\
    
The variations of the IR beam absorption signal appearing in Fig.\ref{absorcion}.b can be the result of MW-induced variations in the Rydberg excitation probability or of variations of the total number of atoms. We argue that the first mechanism can explain the sharp increase of the absorption at MW frequencies corresponding to Rydberg transition frequencies but is unable to account for the above mentioned broad transparency increase. Instead, we claim that the increase in the transparency is the result of the removal of atoms from the cloud as a consequence of the strong DDI between pairs of atoms made possible by the MW field.\\

\begin{figure} [htbp]
\centering
  \includegraphics[width=1\linewidth]{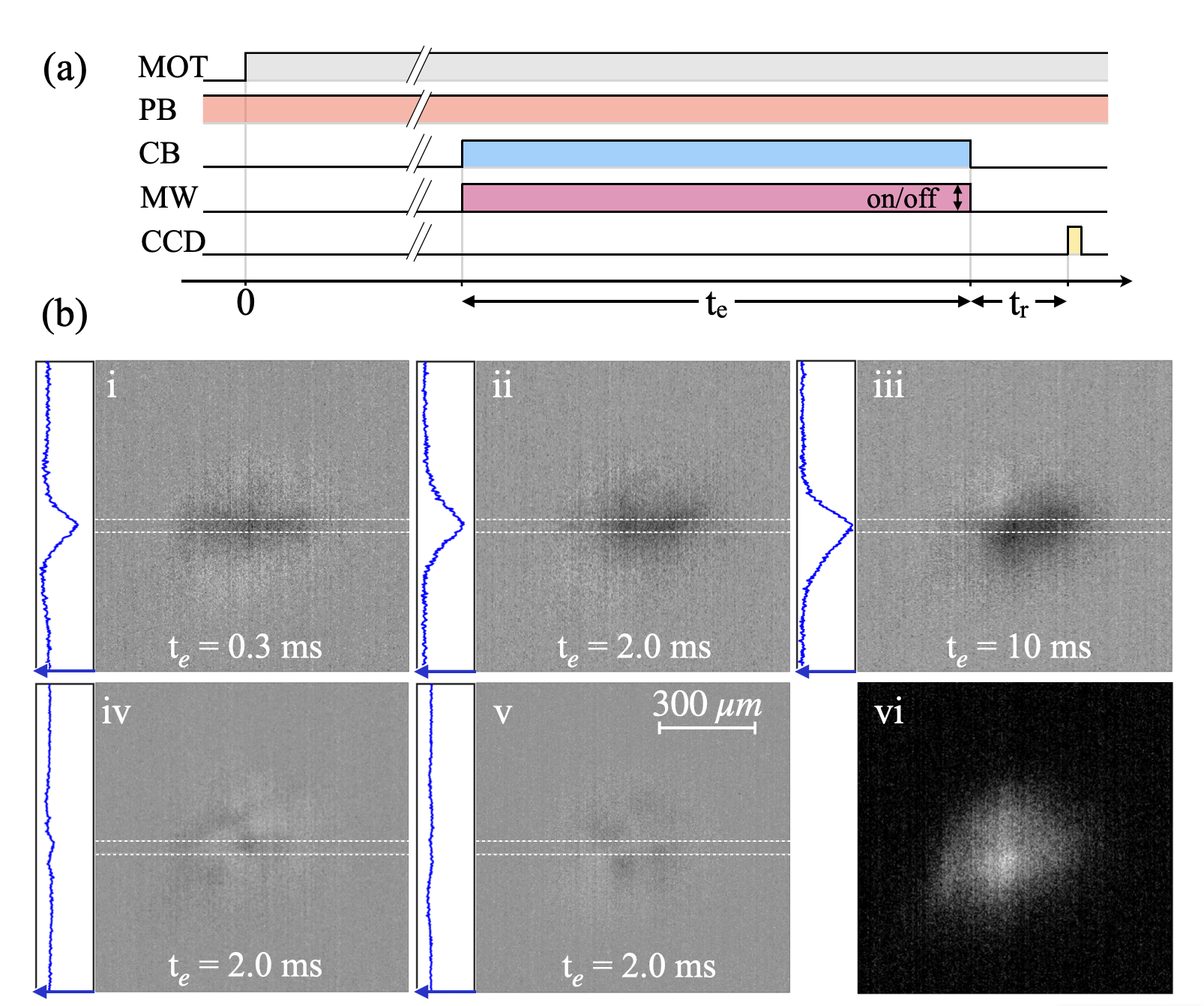}
\caption{\label{imagenes} a) Time sequence (same notation as in Fig. \ref{absorcion}. CCD: trigger pulse to camera).  b) Images of the cold cloud fluorescence after Rydberg optical excitation and alternate MW field. i-v: Difference images between acquisitions with and without MW radiation. i-iii: MW frequency $\simeq 14.5$ GHz and different exposition times $t_e$. iv: MW frequency $\simeq 14.0$ GHz, $t_e =$ 2 ms. v: difference image noise background (without MW field). vi: MOT fluorescence without Rydberg excitation. Left side of images i-v:  fluorescence brightness difference integrated along  beam propagation direction (same intensity scale for all figures). The dashed white lines indicate the approximate path of the Rydberg exciting laser beams across the cold cloud. In i-v the gray background corresponds to zero difference between images and darker tones indicate a negative difference. In vi the black background corresponds to zero fluorescence and lighter tones correspond to increasing fluorescence.}
\end{figure}

To test our claim, we have taken images with a CCD camera of the fluorescence emitted by the cold atomic cloud immediately after turning off the simultaneous irradiation of the cloud by the probe and coupling optical fields and with MW radiation present once every two MOT realizations. Details of the time sequence are given in Fig. \ref{imagenes}.a. The fluorescence was induced by the MOT trapping laser beams. A delay $t_r \sim$ 1 ms was allowed between the turning off of the blue and the MW fields and the acquisition of the image (300 $\mu$s acquisition time). This delay is long enough to allow the decay of excited atoms to the ground state. Under such conditions, the image brightness reflects the number of atoms in the MOT cloud (projected along the line of sight of the camera).\\
The images i-iv in Fig. \ref{imagenes}.b are representations of the \emph{difference} images taken with the MW field present minus the image taken without MW. Image v illustrates the background noise (acquired without the MW field), and image vi was acquired without the excitation of Rydberg atoms to illustrate the cold-cloud atomic distribution and size. Dashed white lines in images i-v indicate the approximate path of the IR and blue Rydberg exciting lasers across the atomic cloud.  

Figures \ref{imagenes}.b.i-iii correspond to different exposure times $t_e$ of the cold cloud to Rydberg excitation and MW radiation (if present). For these images the MW was tuned to 14.5 GHz, corresponding to the minimum absorption in Fig. \ref{absorcion}.b. Image iv was taken with the MW tuned to the position of the absorption peak near the 55D$_{3/2}$ $\rightarrow$ 54F$_{5/2}$ transition frequency ($\sim$14.0 GHz).\\

A reduction of the number of atoms around the region illuminated by the Rydberg exciting laser beams as a result of the MW radiation is visible in images i-iii. We interpret the reduction as removal of atom pairs by the dipole - dipole force induced by the MW. The depleted region grows in directions transverse to the exciting laser beams as the irradiation time with MW radiation increases. This growth is the result of diffusion between the region illuminated by the Rydberg exciting beams and the remaining of the cold cloud. No significant change (above noise level) in the atomic population with and without MW can be seen in image iv.\\  

\begin{figure} [htbp]
\centering
  \includegraphics[width=1\linewidth]{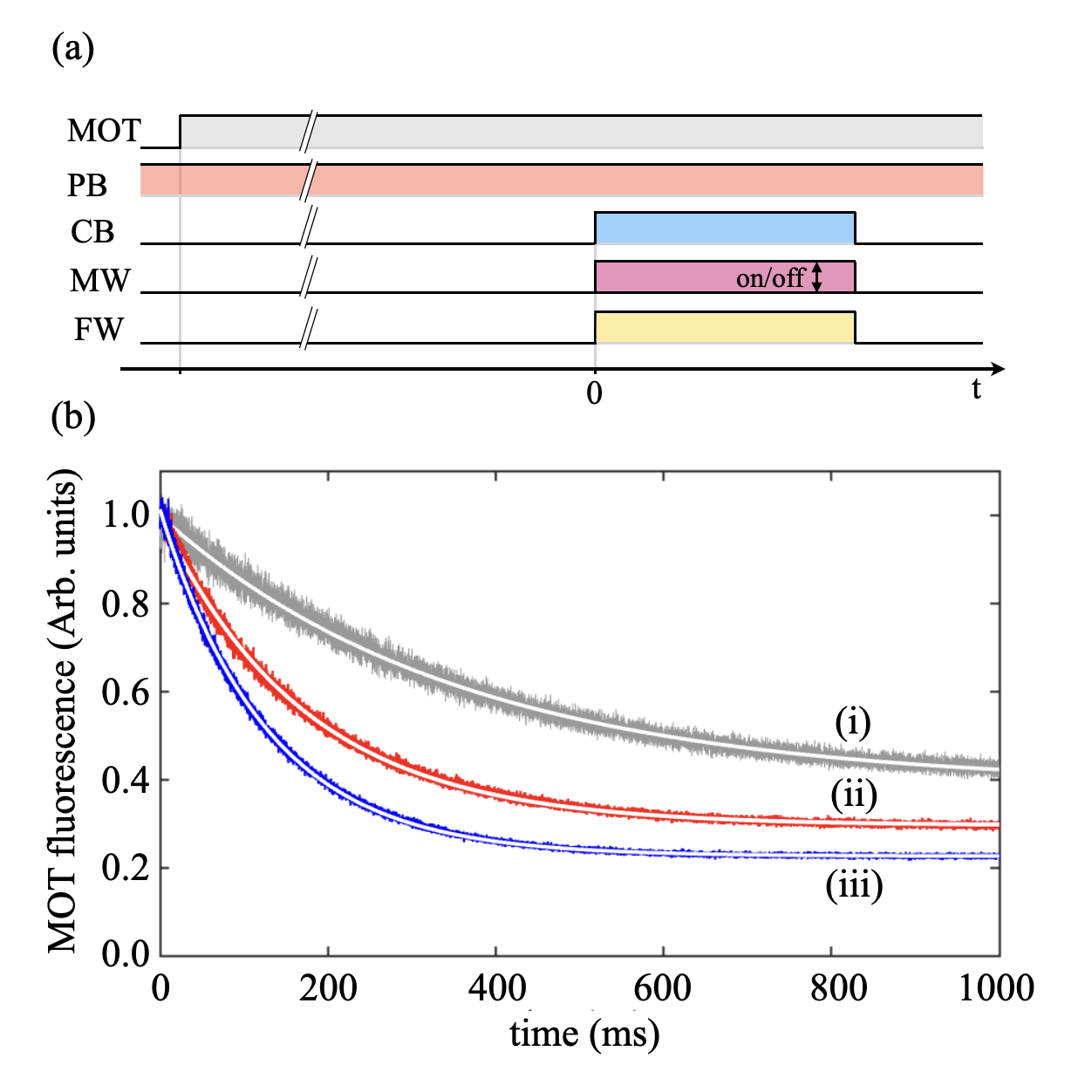}
\caption{\label{transitorios} a) Time sequence (same notation as in Fig.\ref{absorcion}, FW: fluorescence detection interval). b) Transient evolution of the cloud fluorescence i) Blue light   \emph{un}locked to the two-photon transition.   ii) Blue light locked to the Rydberg two-photon excitation resonance.  iii) Same as ii) in the presence of MW radiation (14.5 GHz). The white solid lines are exponential fits.}
\end{figure}

Additional information on the cold-atom loss-mechanisms involved in the experiment can be obtained from Fig.\ref{transitorios}. The figure shows the transient evolution of the atomic cloud fluorescence for a MOT operating under constant conditions (all trapping beams on) in the presence of the IR probe beam, whose effect on the MOT operation is negligible. At $t=0$, after the MOT has been loaded, the blue light is turned on under different conditions. In trace iii the MW field is  present with its frequency tuned to 14.5 GHz, corresponding to the minimum absorption in Fig. \ref{absorcion}.b. 

In trace i the blue light frequency is \emph{not} locked to the Rydberg two-photon transition (detuning $\sim 0.2$ GHz).  The observed loss of atoms presumably results from ionization of atoms in the 5P$_{3/2}$ level by blue photons. The decay rate determined by an exponential fit is $\gamma_{B}=$ (357 ms)$^{-1}$. In traces ii and iii the blue light has been locked to the two-photon resonance to the 55D$_{3/2}$ state. The increased decay rate in trace ii is now $\gamma_{Ry}=$ (173 ms)$^{-1}$ indicating a significant role played by Rydberg atoms in the cold cloud atomic losses. Several mechanisms may be involved in such losses including: free flight and fall under gravity of Rydberg atoms insensitive to the MOT forces, inter-atomic forces between Rydberg atoms due to the vdW interaction or black-body radiation mediated DDI, cold atom collision and ionization \citep{Faoro16,Knuffman06,Park11}.  Trace iii corresponds to a total decay rate  $\gamma_{MW}=$ (127 ms)$^{-1}$ revealing the existence of an additional MW-induced loss mechanism.\\

\begin{figure} [htbp]
\centering
  \includegraphics[width=1\linewidth]{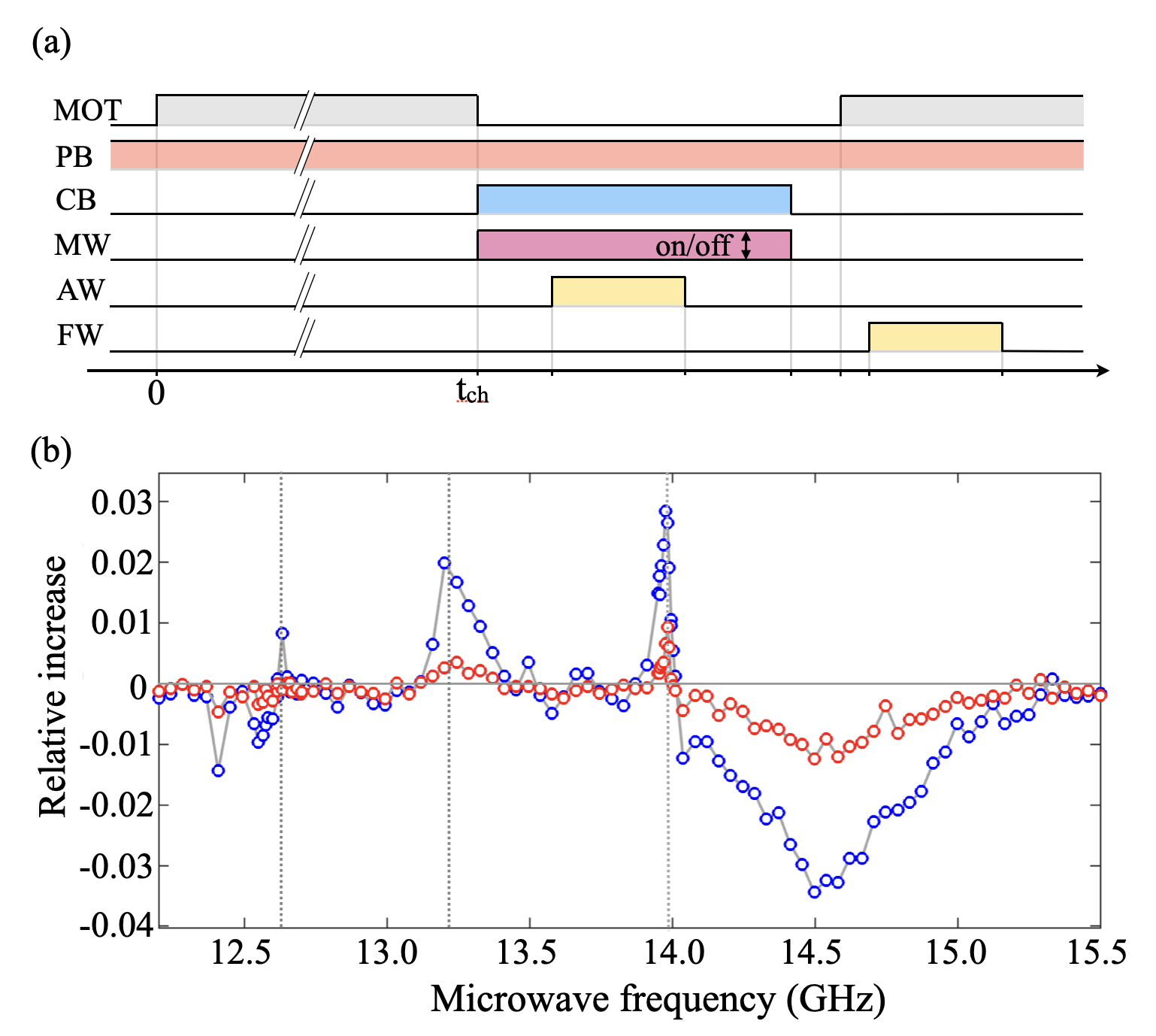}
\caption{\label{absyfluo} a) Time sequence.
b) Relative signal variation [(MW ON signal - MW OFF signal)/ MW OFF signal] as a function of MW frequency. \emph{Blue}: IR beam absorption. \emph{Red}: Cold cloud fluorescence recorded 1 ms after turning off the blue and MW fields. Same dotted lines as in Fig. \ref{absorcion}}
\end{figure}

Figure \ref{absyfluo} presents the change IR absorption and cloud fluorescence, relative to their value in the absence of MW, as a function of the MW frequency recorded sequentially in each cold cloud realization. The cloud fluorescence was recorded 1 ms second after turning off the Rydberg atom excitation by the blue light. This interval is long enough to allow excited atoms to return to the ground state. The fluorescence peak observed near 14 GHz is due to a reduction of the Rydberg excitation probability and consequently a reduction of the Rydberg-atoms-depending atom-loss mechanism operating in Fig. \ref{transitorios}.ii. 
As will be further argued in the theoretical discussion, no significant increase in the Rydberg atom excitation probability is expected for MW frequencies detuned from the transition frequencies between individual-atom Rydberg states. In consequence, the reduction of the fluorescence for MW frequencies above 14 GHz appears as an indication of the reduction of the total number of atoms present in the MOT. Similar patterns are observed in both the absorption and fluorescence spectra in Fig. \ref{absyfluo} strongly suggesting a common origin for the corresponding structures. 

\section{Modeling}

In the  experiments, the MW frequency was scanned over a range including the transition frequencies from states 55D to states 56P and 54F. For simplicity, we will present a model considering only the transitions $55D\ \rightarrow\ 54F$ (in the following we frequently refer to these states as D and F states respectively). As such, our model is not expected to reproduce spectral features for MW frequencies below the D $\rightarrow$ F transition frequency (14.0 GHz), strongly affected by $55D\ \rightarrow\ 56P$ transitions. In addition, the spectra shown in Figs. \ref{absorcion} and \ref{absyfluo} reveal features extending over several hundred MHz to the blue of the D $\rightarrow$ F transition frequency. Such frequency range corresponds to energy shifts larger than the fine structure splittings of 55D and 54F Rb Rydberg levels (69 MHz and 1.0 MHz respectively). We have consequently ignored the spin orbit coupling in the model.\\

Some features present in the spectra cannot originate from the response of independent atoms. We have therefore developed a model for individual atoms and for atom-pairs dressed by a single MW field mode of frequency $\omega$ containing $N$ photons \citep{Cohen98}. In the corresponding dressed states basis we have numerically determined the eigenstates and eigenenergies of the total Hamiltonian including: the atomic Hamiltonian, the MW field Hamiltonian, the atom (or atom-pair) interaction with the MW and in the case of atom pairs, the DDI. The transition amplitudes and probabilities for Rydberg excitation by the optical fields of dressed single-atoms or dressed atom-pairs eigenstates were calculated using first-order or second-order perturbation theory respectively. The details of the calculation are provided in the Appendix. \\

\begin{figure}[ht!]
\centering
\includegraphics[width=1\linewidth]{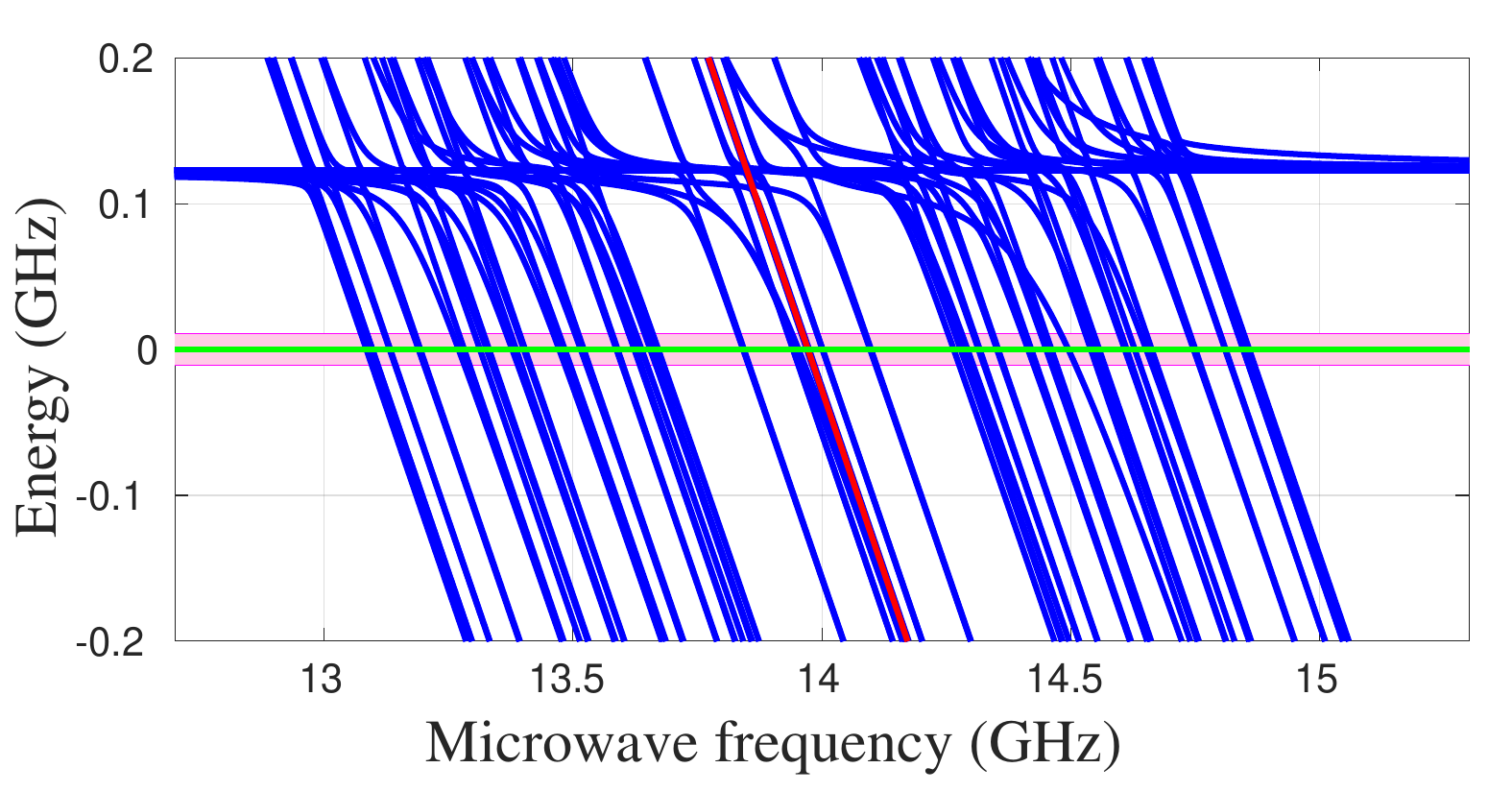}
\caption{\label{acoplados} Energies of the dressed Rydberg atom-pairs states in the rotating frame evolving at frequency $N\omega$ as a function of the MW frequency. Green: energy of the \emph{uncoupled} states $\ket{D,D,N}$ taken as zero. Red: energy of the \emph{uncoupled} states $\ket{D,F,N-1}(\ket{F,D,N-1})$ Blue: eigen-energies of the total Hamiltonian $H$. The spreading of the parallel asymptotic lines is due to the dipole - dipole coupling between $\ket{D,F,N-1}$ and $\ket{F,D,N-1}$ states. The avoided crossings are the consequence of the coupling to the MW field. The horizontal pink-colored area represents the optical excitation bandwidth ($\pm\ \kappa$). [Parameters.  Inter-atomic distance: $R= 2.2\ \mu m$. Polar angle: $\theta = \pi/2$. Microwave Rabi frequency: $\Omega_{MW}= 2\pi \times 43\ MHz$. Decoherence rate: $\kappa= 2\pi \times 11\ MHz$.]}
\end{figure}

\begin{figure}[ht!]
\centering
\includegraphics[width=1\linewidth]{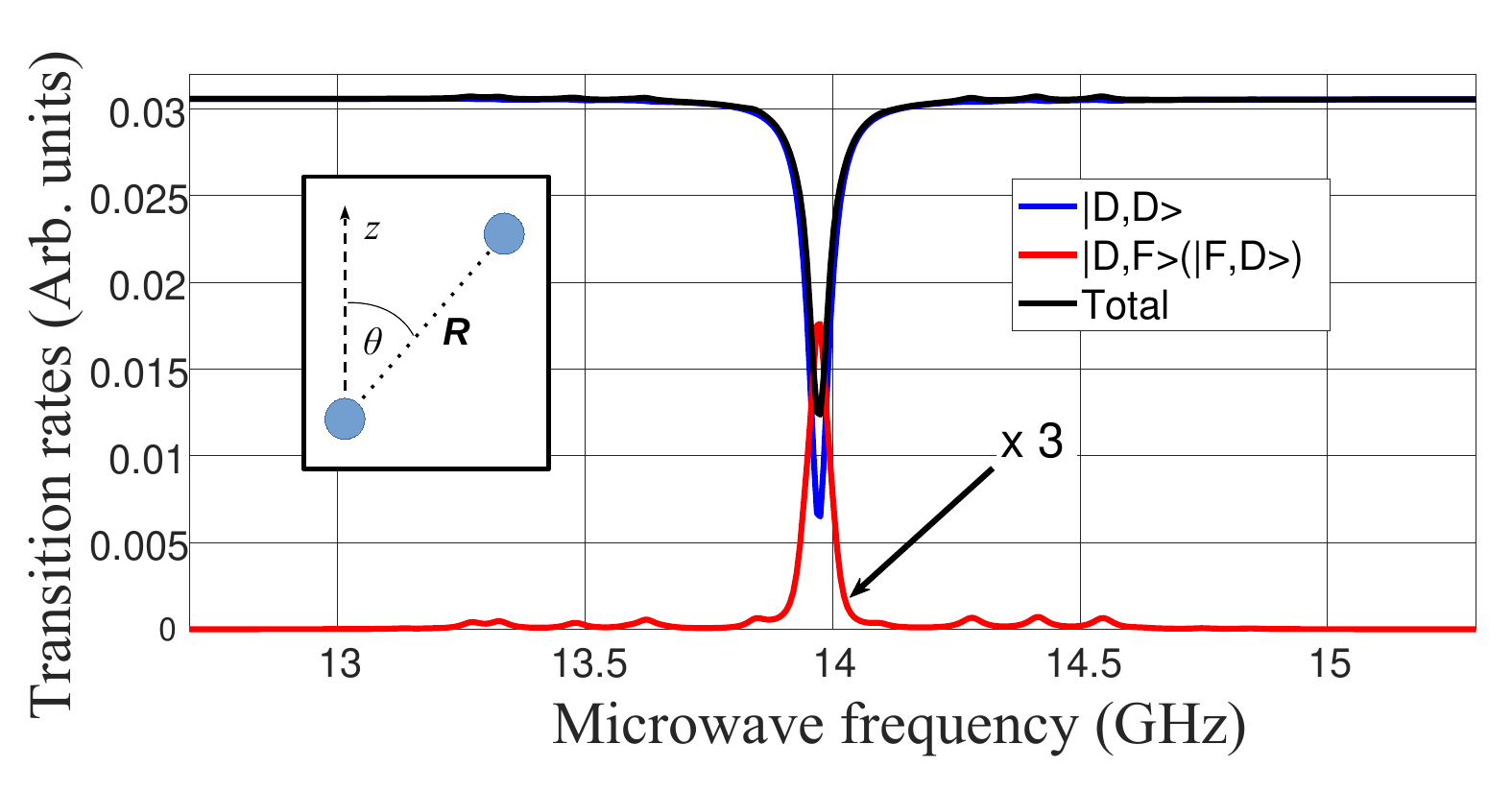}
\caption{\label{probabilidades} Excitation probability for double Rydberg pair excitation for the conditions mentioned in Fig. \ref{acoplados}. Black: total probability. Blue: $\ket{D,D}$ states excitation probability. Red: $\ket{D,F}$ or $\ket{F,D}$ excitation probability (multiplied by 3). Inset: atom-pair geometry relative to the quantization axis $z$.}
\end{figure}

Figure \ref{acoplados} provides an example of the energy eigenvalues of the total Hamiltonian for a specific atom-pair (corresponding parameters indicated in the figure caption) represented in a reference frame rotating at frequency $N\omega$. The \emph{uncoupled} energies are the straight red and green lines crossing at the 55D $\rightarrow$ 54F transition frequency. The coupling between the atoms and the MW is responsible for the avoided crossings while the DDI is responsible for the lifting of the degeneracies resulting in many parallel asymptotic lines. The displacement above zero ($\sim$ 0.13 GHz) of the horizontal asymptotic lines is the consequence of the vdW interaction within $\ket{D,D,N}$ states assumed to be isotropic (The coefficient $C_6$ = 15 GHz.$\mu m^6$ was used \citep{Reinhard07}).  The total number of energy curves is mainly dependent on the dimension of the considered Hilbert space basis. Most notable, the avoided crossings extend on each side of the single-atom transition frequency over a broad MW frequency range. Their actual positions depend on the inter-atomic distance and the polar angle between the inter-atomic axis and the quantization axis (see inset in Fig.\ref{probabilidades}), however the trend illustrated in Fig. \ref{acoplados} is general.\\

The order of magnitude of the energy shifts due to DDI shown in Fig. \ref{acoplados} deserves a comment. The DDI Hamiltonian (Eq. \eqref{H_DD}) can be factored into the energy $\xi=\mu^2/(4\pi\epsilon_0R^3)$ multiplied by a numerical matrix to be evaluated by standard angular momentum algebra ($\mu$ is the reduced electric dipole matrix element for the transition). For the conditions of Fig.\ref{acoplados} $\xi \simeq h \times 4.1$ GHz and the largest eigenvalue of the numerical matrix is approximately 0.2. Accordingly, the spread of energies shifts extends over a range of approximately $\pm$ 0.8 GHz around the $D \rightarrow F$ transition frequency.\\   

Considering that the optical fields only couple the atomic ground state to the 55D$_{3/2}$ level, 54F states can only be excited near the avoided crossings where mixing between D and F states occur. The excitation probability of dressed atom-pair eigenstates can be calculated using second order perturbation theory with singly-excited atom pairs as intermediate levels. The result of the calculation for the atom-pair considered in Fig. \ref{acoplados} is shown in Fig. \ref{probabilidades}. The main feature of this plot which is independent of the specific atom-pair geometry, is the minimum total excitation probability occurring at the D $\rightarrow$ F transition frequency (black curve). It is the consequence of the reduction of the probability of single-excitation (intermediate state) due to the avoided crossing [Autler-Townes (AT) effect] resulting from the interaction with the MW \cite{Tanasittikosol11,Brekke12}. The reduction of the double Rydberg excitation probability of the atom-pair (blue curve) is only partially compensated by the increase in the probability of exciting atoms in $\ket{D,F}$ or $\ket{F,D}$ states (red curve). However, such probability-increase extends over a broad frequency range corresponding to the multiple avoided crossings in Fig. \ref{acoplados}.\\

The dependence on MW frequency of the double Rydberg excitation probability of atom-pairs presented in Fig. \ref{probabilidades}, although corresponding to a particular example, is representative of the structure obtained for any pair geometry. In an ensemble of randomly distributed pairs as encountered in the cold atom cloud, all pairs contribute to the reduction of the total double excitation probability observed at the D $\rightarrow$ F transition frequency and no significant increase of the total excitation probability is obtained for any MW frequency. We conclude that the positive and negative variations of the IR absorption with MW frequency observed in Figs. \ref{absorcion} and \ref{absyfluo} cannot be explained both as resulting only from variations of the excitation probability.

However, since the IR absorption is proportional to the ground state population, a reduction of the absorption can be explained by the depletion of the number of atoms probed by the IR beam. Also, we notice that the broad MW frequency range where the reduction of the IR absorption occurs, roughly corresponds to the frequency range where the probability excitation of atom pairs in $\ket{D,F}$ or $\ket{F,D}$ states is increased (see the red curve in Fig. \ref{probabilidades}). Atoms in such pair states are sensitive to DDI and experience an inter-atomic force sufficiently strong to expel the atoms from the observed volume in a few tenths of microseconds (see Fig. \ref{dxt} in Appendix).\\  

In order to connect the calculated dressed atom-pairs excitation probabilities and the atomic depletion hypothesis with the observed spectra, a model of the atom-cloud dynamics is required. Such a model should include the dynamical evolution of the internal and external degrees of freedom of the atomic ensemble. Moreover, to be accurate, the model should also account for some specificities of our experiment such as the need to turn off the MOT trapping beams (to avoid ionization) during the IR absorption measurements resulting in the free fall of the cold cloud and consequently a limitation of the observation time. Also in our setup the Rydberg-exciting-beams cross-sections are smaller than the cold-cloud dimension (see Fig. \ref{imagenes}). Therefore, the total atomic ensemble can be divided into two fractions, one that interacts with the exciting light and one that does not. Diffusive motion between these two fractions further complicates the atomic dynamics. 

\begin{figure}[ht!]
\centering
\includegraphics[width=1\linewidth]{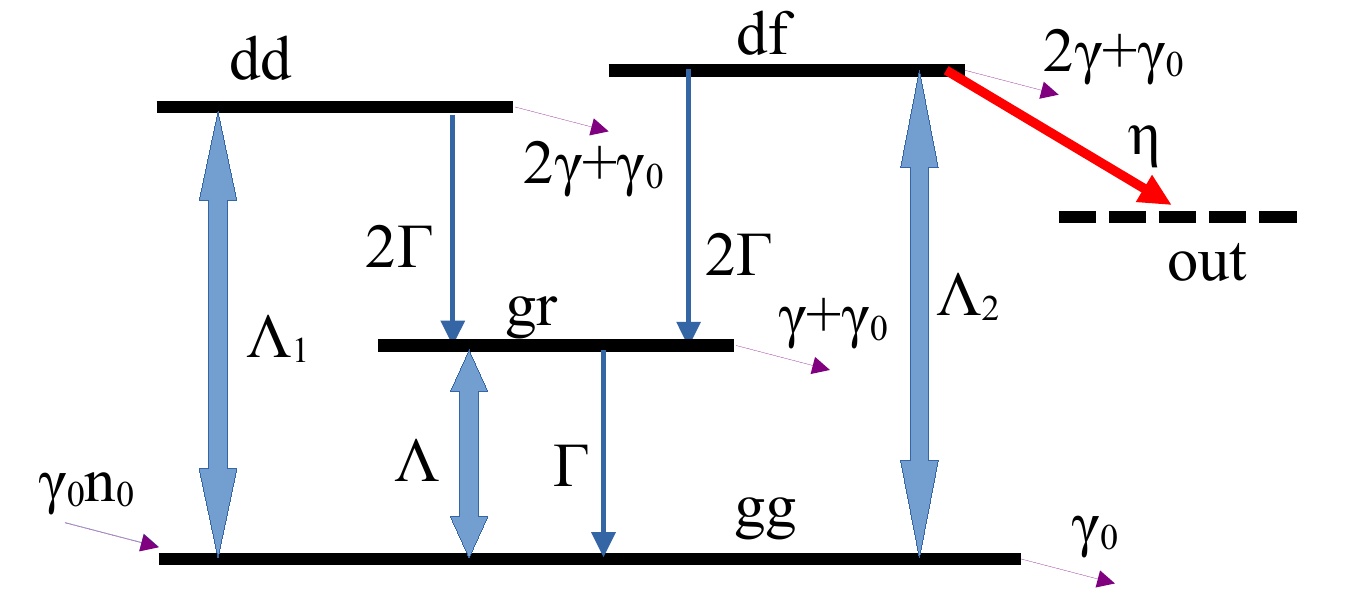}
\caption{\label{singledouble} Simplified dressed atom-pair level scheme including single a doubly excited states.}
\end{figure}

In view of these difficulties, we have refrained from modeling the actual experimental conditions. Instead, we present a simple  rate equations model that includes a minimal set of assumptions. The model refers to the level scheme presented in Fig. \ref{singledouble}.\\ 

The model is concerned with a homogeneous ensemble of atom-pairs  with fixed inter-atomic distance $R$ and polar angle $\theta$.  A simplified notation is used to designate the pair dressed-state. $gg$ refers to the pair ground-state (MW photon numbers are implicit). $gr$ refers to pair states with only one excited Rydberg atom. $dd$ and $df$ refer to doubly excited Rydberg pairs. $dd$ includes all Rydberg states not affected by the dipole - dipole repulsion [mainly $\ket{D,D}$ states and some $\ket{F,D}(\ket{D,F})$ states for which the DDI is not sufficiently large (see Eq. (\ref{expulsion}.a)]. $df$ includes all $\ket{F,D}(\ket{D,F})$ states for which the DDI is significant (Eq. (\ref{expulsion}.b). All level degeneracies are ignored. \\

The model assumes that the ensemble of atom-pairs is fed by atom-pairs in the $gg$ state at a rate $\gamma_0 n_0$. $\gamma_0$ is the intrinsic cold cloud loss-rate. In the absence of any additional loss mechanism, the total atomic population is $n_0$. The decay to the ground state of a Rydberg atom  occurs at rate $\Gamma$ \citep{Gounand79,Saffman05}. Based on the observations presented  in Fig. \ref{transitorios}, we assume a loss-rate $\gamma$ for all Rydberg states ($\gamma > \gamma_0$). In addition, to account for the MW induced dipole - dipole atom removal we include an escape rate $\eta$ from level $df$ out of the pairs ensemble. The excitation rate from $gg$ to $gr$ is designated as $\Lambda$ while $\Lambda_1$ and $\Lambda_2$ correspond to the excitation rates from $gg$ to $dd$ and $df$ respectively.\\

The single-excitation rate $\Lambda$ is calculated from  transition amplitudes obtained using first order perturbation theory while the double excitation rates $\Lambda_1$ and $\Lambda_2$ are calculated using second order perturbation theory. The discrimination between $\Lambda_1$ and $\Lambda_2$ is based on the dipole - dipole potential energy associated to the doubly excited pair-state. The excitation rates calculations assume a ground to Rydberg state dephasing rate $\kappa$. Additional details are provided in the Appendix.\\

The corresponding rate equations are:

\begin{subequations} \label{ecuacionesdetasa_single_double}
\begin{align}
\dot{n}_{gg} =& \gamma_0 n_0-(n_{gg}-n_{gr})\Lambda-(n_{gg}-n_{dd})\Lambda_1\nonumber\\
& -(n_{gg}-n_{df})\Lambda_2+\Gamma n_{gr} -\gamma_0 n_{gg} \label{gg} \\
\dot{n}_{gr} =& (n_{gg}-n_{gr})\Lambda+2\Gamma (n_{dd}+n_{df})\nonumber\\
&-(\Gamma+\gamma+\gamma_0)n_{gr}\label{gd}\\
\dot{n}_{dd} =& (n_{gg}-n_{dd})\Lambda_1-(2\Gamma+2\gamma+\gamma_0) n_{dd} \label{dd}\\
\dot{n}_{df} =& (n_{gg}-n_{df})\Lambda_2-(\eta+2\Gamma +2\gamma+\gamma_0) n_{df} \label{df}
\end{align}
\end{subequations}
\newline

The total number of parameters involved in the calculation of the excitation probabilities and in the rate equations \eqref{ecuacionesdetasa_single_double} is very large. It is impractical to independently vary those parameter to fit the data. Some of these parameters can be accessed from the literature while others need to be estimated from the experimental conditions. We have not attempted to fully determine all parameters values but to show that a reasonable agreement between experimental observations and modeling can be reached for reasonable values of the parameters. In that spirit, the parameters $\Gamma$, $\gamma$, $\gamma_0$, $E_{th}$ and $\Omega_{MW}$ were just given order of magnitude values compatible with previous literature or experimental estimates.

In addition to the parameters mentioned above, the modeling depends on the values of the effective optical excitation Rabi frequency $\Omega_{Opt}$ and the two-photon transition decoherence rate $\kappa$ which were given reasonable values estimated from the involved laser characteristics. 

Finally, parameters $\eta$ and the typical interatomic distance in atomic pairs $a$ were varied for  qualitative agreement between observations and numerical simulation. The actual values used for the parameters are given in Table \ref{valoresparametros} in the Appendix.\\

\begin{figure}[ht!]
\centering
\includegraphics[width=1\linewidth]{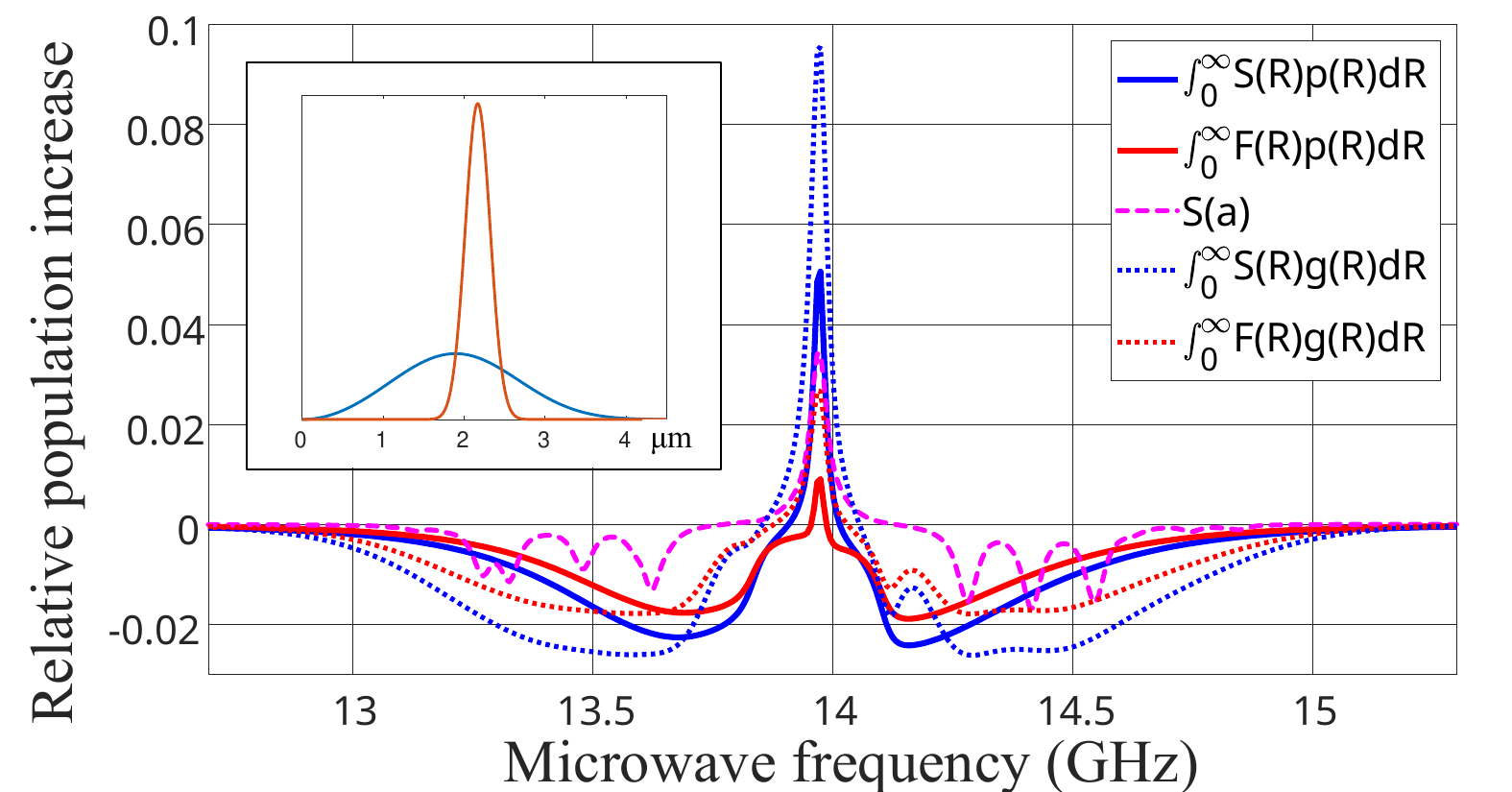}
\caption{\label{espectroteorico} Calculated relative atom-number variation taking into account the inter-atomic distance distribution and the polar angle dependence of the DDI. Blue: ground state population. Red: all levels population. Solid lines: inter-atomic distance distribution $p(R)= \frac{3}{a}(\frac{R}{a})^2\exp{[-(\frac{R}{a})^3]}$ \citep{Hertz1909,Chandrasekhar43,Santos86}. Dotted lines:  Gaussian interatomic distance distribution $g(R) \propto \exp{[-(\frac{R-a}{w_a})^2]}$ with $w_a=a/10$.  The dashed magenta line (not to scale) corresponds to the ground-state population variation for a fixed inter-atomic distance $a = 2.2\ \mu$m and polar angle $\theta=0$. In the legend $S(R)$ and $F(R)$ refer to the polar-angle-averaged ground state population and total population respectively for a given inter-atomic distance $R$. Inset: Pair distance distributions $p(R)$ (blue) and $g(R)$ (red). The values of the parameters used in the simulation are given in Table \ref{valoresparametros} in the Appendix.}
\end{figure}

The steady-state solution of Eqs. \eqref{ecuacionesdetasa_single_double} was computed  and used to evaluate two quantities: the ground-state population $n_{gg}$ assumed to be the main parameter controlling the IR laser absorption and the total atomic population determining the MOT fluorescence brightness. As an example, the ground state population obtained for fixed inter-atomic distance $R=a=\ 2.2\ \mu$m and polar angle $\theta = \pi/2$ is represented by the dashed magenta line in Fig. \ref{espectroteorico}.\\ 

In order to relate the theoretical results with the experimental observation, a pair distance distribution has to be assumed in the cold sample. We consider a frozen sample approximation \citep{Teixeira15,Park16} with an isotropic distribution of the orientation of the inter-atomic axis and an inter-atomic distance distribution corresponding to the nearest-neighbor distance distribution in an ideal gas of point-like classical particles: $p(R)= \frac{3}{a}(\frac{R}{a})^2\exp{[-(\frac{R}{a})^3]}$ \citep{Hertz1909,Chandrasekhar43,Santos86}
For comparison, we also consider a more localized Gaussian interatomic distribution $g(R) \propto \exp{[-(\frac{R-a}{w_a})^2]}$ with $w_a=a/10$ (see inset in Fig. \ref{espectroteorico}).\\

The blue traces in Fig. \ref{espectroteorico} show the relative change in the \emph{ground state} population (representative of the IR absorption) obtained after integration over $R$ and $\theta$ using the two pair-distance distributions $p(R)$ and $g(R)$. The red traces correspond to the variation of the \emph{total} atomic population (MOT brightness). The parameters used in the calculations are indicated in the figure caption. The values of the parameters $\kappa, \ \Omega_{Opt},\ \Omega_{MW}$ and $\Gamma$  are representative of the actual experimental conditions. The rates $\gamma_0$, $\gamma$ and $\eta $ were estimated from the transient signal evolutions presented in Fig. \ref{transitorios}. \\

The theoretical modeling relies on several simplifying assumptions among which: a) Small number of basis states limited to Rydberg levels 55D and 54F of Rb. b) Neglect of the spin orbit interaction (preventing the inclusion of states 56P in the basis set). c) Simplistic treatment of the vdW interaction for pairs of Rydberg atoms in $\ket{D,D}$ states considered isotropic.  This is a consequence of the reduced number of states included in the basis set [assumption a)]. d) Frozen cloud approximation. e) Ensemble of independent pairs. No many body effect was considered. f) Simplistic inter-atomic distance distribution for atom pairs. g) Rate equation treatment of the atomic evolution. h) Constant decay rates independent of the pair geometry and magnetic sublevels. i) Steady state solution of the rate equations system.

Nevertheless, in spite of its many simplifying assumptions, the model  succeeds in reproducing the essential features of the spectroscopic observations. It correctly reproduces the absorption/MOT-brightness  peaks observed for MW frequencies resonant  with Rydberg transitions which are the consequence of the the AT effect. It also predicts the increase in transparency/atom-loss over a broad MW frequency range corresponding from the lifting of the $\{\ket{D,F},\ket{D,F}\}$ dressed atom-pair energy degeneracy by the DDI.\\

\section{Discussion} \label{discusion}

The spectrum presented in Fig.  \ref{absorcion} reveals a significant increase in the sample transparency when the MW frequency is tuned over a broad frequency range to the blue of the D $\rightarrow$ F transition. Maximum transparency occurs around 14.5 GHz. At this frequency the EIT increase is comparable in size to the EIT reduction observed on resonance with the 55D $\rightarrow$ 54F transition. The images presented in Fig. \ref{imagenes} confirm that the increased transparency, occurring when a MW frequency around 14.5 GHz is used, results from a depletion of the atomic cloud along the path of the Rydberg exciting lasers beams. The transient evolutions shown in Fig. \ref{transitorios} reveal the existence of an additional atom-loss channel allowed by the presence of the MW. Finally, the trap-loss spectrum plotted in Fig. \ref{absyfluo} together with the absorption spectrum clearly points to the variation of the number of atoms in the cloud as the common cause of the observed spectral features.\\

It is well established that atomic excitation into Rydberg levels results in the loss of atoms from the cold atomic cloud  \citep{deHond20,Cao22,Halter23,Duverger24,Kondo24}. In consequence, the atom-number reduction responsible for the increased transparency occurring for MW frequencies around 14.5 GHz could, in principle, be explained by an increase of the Rydberg excitation probability. However, as illustrated in Fig. \ref{probabilidades}, only minute enhancements of the double pair excitation probability were obtained for all the tested sets of parameter values consistent with the experimental conditions. For a given choice of the pair geometry, the small enhancements of the transition probability observed on either side of the central peak are roughly two orders of magnitude smaller than the central reduction of the excitation probability. The relative magnitude of the central excitation probability-reduction compared to the off-resonance probability-increase structures becomes even larger after radial and angular integration since the central structure is always present while the small off-resonant structures are displaced depending on $R$ and $\theta$. \\

Given the quantitative failure of the transition probability variations to explain the observed atom-loss occurring for MW frequencies detuned from resonance, we notice that for such MW frequencies, $\ket{D,F}(\ket{F,D})$ pairs can be excited as a consequence of the DDI (red trace in Fig. \ref{probabilidades}). The DDI potential energy of the doubly excited pairs roughly corresponds to the absolute value of the MW detuning from the Rydberg transition frequency (several hundred MHz). The resulting force is sufficient to accelerate the atoms and expel them from the observed region in a few tenths microseconds (see Fig \ref{dxt} in the Appendix). We hypothesize that  such mechanical effect is at the origin of the observed  MW-induced atom removal. The rate equation model developed above based in our hypothesis leads to predicted spectra in good qualitative agreement with the observations. \\ 

It is interesting to observe that the agreement between the observations and the calculated spectra appears to be somehow better when the narrower Gaussian radial probability distribution $g(R)$ is used (dotted lines in Fig. \ref{espectroteorico}) instead of the  nearest neighbor distribution $p(R)$. This may indicate that the atomic positions in the cloud may not actually be random or that many-particle effect could play a role. In fact, regular spatial arrangements or the emergence of spatial correlations have been observed or predicted in strongly interacting Rydberg atoms (under conditions different from those of our experiment) \citep{Glaetzle12,Garttner13,Schauss12}. 

\section{Conclusions}

We have presented experimental observations that reveal the existence of a MW induced atom-loss mechanism occurring for MW frequencies detuned from Rydberg transition frequencies by several hundred MHz. The experimental results as well as the theoretical model provide strong indication that the MW induced atom removal is the result of the mechanical effect arising from strong DDI between doubly excited Rydberg atom pairs.\\

The MW induced mechanical effect revealed here can be considered as a practical means to manipulate Rydberg atom systems. Ideally, it could allow the removal of selected Rydberg pairs.
Unfortunately, one must keep in mind that the selection of the MW frequency does not warrant spatial selectivity given the fact that different pair geometries have similar dressed-pair energies (see Fig. \ref{acoplados}). However, the number of the DDI eigenvalues depends on the number of magnetic sublevels involved in the transition (12 for a D $\rightarrow$ F transition), the spatial selectivity should be substantially improved for an S $\rightarrow$ P transition.\\

The experimental observations presented in this article, as well as their interpretation, are largely affected by the disordered nature of our sample (cold gas). A better insight into the mechanisms at work and their application is to be expected from experiments performed on better-controlled atomic samples such as dipolar traps \citep{Kurdak25} or optical lattices \citep{Browaeys20}. 

\section{Acknowledgments}
The authors are thankful to L. Lenci for his contribution to the early stages of the experiment, to A. Saez for valuable technical assistance and to C. Cormick for pertinent remarks on the manuscript. This work was supported by ANII, CSIC and PEDECIBA (Uruguayan agencies). 

The data that support the findings of this article are openly available \citep{Repository25}.

\appendix*
\section{Dressed states model}
\subsection{Dressed single Rydberg atom}
\subsubsection{Basis states}

Considering a microwave field whose frequency is close to the transition-frequency between atomic levels 55D and 54F ($\sim 14$ GHz), the  basis states considered for dressed individual Rydberg atoms are the \emph{quasi-degenerate} set of direct products of normalized atomic states and single-mode microwave field states \citep{Cohen98}:
\begin{eqnarray*}
\mathcal{B}^{(1)}(N) &=& \left\lbrace \ket{55D_{\alpha}}\ket{N},\ket{54F_{\alpha^{\prime}}}\ket{N-1} \right\rbrace 
\end{eqnarray*}
where $\alpha(\alpha^{\prime})$ designates additional atomic quantum numbers. $N$ is the photon number of the dressing microwave field of frequency $\omega$. Assuming that the MW field is in an intense single mode coherent  state ($\Delta \hat{n} = \braket{\hat{n}}^{1/2}\ll \braket{\hat{n}}$), we use $N = \braket{\hat{n}}$,  $\hat{n}$   being the photon number operator \citep{Cohen98}.

We signal that atomic states whose energy differences with the 55D states are comparable to integer multiples of $\hbar \omega$ are not included in the basis although such states may exist in the close-packed state-density characteristic of Rydberg atoms. \\ 

In the following, we will frequently use a simplified notation for the basis states in which the principal quantum number, the additional atomic state numbers $\alpha$ and the microwave photon number will be implicit. \\

\subsubsection{Hamiltonian}

In the Hilbert space sustained by $\mathcal{B}^{(1)}(N) $ and in the rotating frame evolving at frequency $N\omega$ the dressed atom Hamiltonian is:

\begin{eqnarray*}
H^{(1)} &=& H_0^{(1)}+H_{MW}^{(1)}
\end{eqnarray*}

Where, 
\begin{eqnarray}
H_0^{(1)} &=& \hbar (\omega_{FD}-\omega)\sum_{\alpha} \ket{F_{\alpha}}\bra{F_{\alpha}}
\end{eqnarray}
$\hbar\omega_{FD}\equiv \hbar(\omega_{F}-\omega_{D})$ is the energy difference between states 54F and 55D. The energy of the uncoupled dressed states $\ket{D,N}$ is taken as zero. The uncoupled energies group into two branches, one of them with zero energy and the other linearly dependent on $\omega$ with slope -1. The two branches intercept at $\omega=\omega_{FD}$.\\

The coupling with the microwave field is described by:
\begin{eqnarray}
H_{MW}^{(1)} &=& \hbar \Omega_{MW}\mathbf{\hat{z}}\
\end{eqnarray}
here $\Omega_{MW}\equiv \mu \xi N^{1/2}/\hbar$ is the microwave Rabi frequency ($\mu = \braket{55D||D||54F}$ is the electric dipole reduced matrix element for the transition, $\xi$ is the single-photon electric field amplitude for the MW mode). $\mathbf{\hat{z}}$ is a dimensionless vector operator aligned with the microwave linear polarization which is chosen as the quantization axis ($\mathbf{\hat{z}}\equiv T^{(1)}_0$  where $T^{(1)}$ is a spherical tensor operator). 

$H_{MW}^{(1)}$ is responsible for the avoided crossing of the two energies branches and the mixing by the microwave of $\ket{D}$ and $\ket{F}$ states. \\ 

\subsubsection{Optical excitation}

The optical fields are responsible for the excitation of states within the basis set $\mathcal{B}^{(1)}(N)$ from the dressed ground states $\{\ket{5S_{1/2},N}\}$ set. The MW field-state $N$ is not affected by the optical interaction.\\ 

Guided by the experimental conditions, we assume that all optical fields involved in Rydberg levels excitations are linearly polarized along the quantization axis $z$. This assumption imply the conservation of the quantum number $m_J$ (total angular momentum component along the quantization axis).\\ 

We effectively describe the optical excitation as equivalent to the irradiation of the atoms with a monochromatic optical wave of frequency $\omega^{\prime}$ and Rabi frequency  $\Omega_{Opt}$. In the rotating-wave approximation, the corresponding interaction is described by:
\begin{eqnarray}
V_{Opt}^{(1)} &=& \hbar \Omega_{Opt}(\ket{{55}D_{\alpha_0}}\bra{g}e^{-i\omega^{\prime}t}\nonumber\\
&& +\ket{g}\bra{55D_{\alpha_0}}e^{i\omega^{\prime}t})\otimes\ \mathds{1}_N\label{Vopt}
\end{eqnarray}
with $\ket{g}$ designating the atomic ground state ($5S_{1/2},m_J=1/2$) and $\alpha_0$ representing the quantum numbers $\{J=3/2,m_J=1/2\}$. $\mathds{1}_N$ is the unit operator for MW field states. 

Only one magnetic quantum number, $m_J=1/2$, was included in \eqref{Vopt}. The contribution of atoms in the $m_J=-1/2$ states can be considered independently in a similar way. 

\subsection{Dressed Rydberg atom-pairs }

\subsubsection{Basis states}

We consider a pair of Rydberg atoms labeled 1 and 2. The basis states set for dressed Rydberg pairs are the products of single-atoms and single-mode field states:

\begin{align}
\mathcal{B}^{(2)}(N) &= \left\lbrace \ket{55D_{\alpha}}\ket{55D_{\alpha^{\prime}}}\ket{N},\right.\nonumber \\
&  \ket{55D_{\alpha}}\ket{54F_{\alpha^{\prime}}}\ket{N-1},\nonumber \\
& \left. \ket{54F_{\alpha}}\ket{55D_{\alpha^{\prime}}}\ket{N-1} \right\rbrace 
\end{align}
where the first ket in the product refers to atom 1 and the second to atom 2. \\

The restriction of basis $\mathcal{B}^{(2)}(N)$ to such a reduced number of states is intended to limit the size of the matrices used in the numerical computations. It has however important consequences. The basis only includes states that can be excited involving at most one MW photon. Higher order processes regarding the interaction with the MW are therefore not accounted for. Also, the basis does not include atom-pair states such as $\ket{P,F}$ whose energy difference  with states $\ket{D,D}$ (F\"{o}rster defect) can be small. This prevents our simplified model to account for van der Waals interaction between atoms in $\ket{D,D}$ states. Such interaction is responsible for the well known Rydberg blockade \citep{Saffman10}. In the present case, instead of increasing the dimensionality of the Hilbert space, the van der Waals interaction was included phenomenologically as described below.\\ 

In the modeling we are brought to consider Rydberg atom-pairs for which the DDI is stronger than the Spin-Orbit interaction (the fine structure splitting of the 55D and 54F are 69 MHz and 1.0 MHz respectively). We will therefore neglect in the calculation the Spin-Orbit interaction and characterize the atomic state with the quantum  numbers $\alpha\equiv \left\lbrace m_L,S,m_S \right\rbrace $ where $S=1/2$ and $m_L$ and $m_S$ are the magnetic orbital and spin quantum numbers. We observe that neglecting the Spin-Orbit coupling would be a poor approximation if the 56P states are included in the basis. For these states the fine structure splitting is 572 MHz, comparable to DDI energies considered here.\\

In the rotating frame evolving at frequency $N\omega$, we take as zero the energy of the degenerate basis states $\ket{55D_{\alpha},55D_{\alpha^{\prime}},N}$. The energies considered in the $\mathcal{B}^{(2)}(N)$ basis set are typically larger than that of the  $\mathcal{B}^{(1)}(N)$ set by an amount of the order of the excitation energy of 55D Rydberg atoms. The two sets can only be coupled via optical interaction.\\

\subsubsection{Hamiltonian}

The atom-pair+field Hamiltonian restricted to the Hilbert space sustained by basis $\mathcal{B}^{(2)}(N)$ is:

\begin{eqnarray*}
H^{(2)} &=& H_0^{(2)}+H_{MW}^{(2)}+H_{DD}+H_{vdW}
\end{eqnarray*}

Where, 
\begin{eqnarray}
H_0^{(2)} &=& \hbar (\omega_{FD}-\omega)\nonumber\\
&\times&( \mathds{1}-\sum_{\alpha,\alpha^{\prime}} \ket{D_{\alpha},D_{\alpha^{\prime}}}\bra{D_{\alpha},D_{\alpha^{\prime}}})
\end{eqnarray}
The eigenvalues of $H_0^{(2)}$ are represented by the green and red lines in Fig. \ref{acoplados}).\\

The DDI Hamiltonian is given by: 
\begin{eqnarray} \label{H_DD}
H_{DD} &=& \frac{e^2}{4\pi\varepsilon_0 R^3}\left[ \mathbf{r_1}\cdot \mathbf{r_2}-3\frac{(\mathbf{r_1}\cdot \mathbf{R})(\mathbf{r_2}\cdot \mathbf{R})}{R^2}\right]
\end{eqnarray}
here $e$ is the elementary charge, $\varepsilon_0$ the vacuum permitivity, $\mathbf{R}$ is the inter-atomic separation vector and $\mathbf{r}_i$ are the electron position vectors relative to the corresponding atomic nucleus. The Hamiltonian $H_{DD}$ includes the dependence with the polar angle $\theta$ between the inter-atomic axis and the quantization axis $z$ \citep{Vermersch15}.  The details of the calculation of the matrix elements of $H_{DD}$ in the basis $\mathcal{B}^{(2)}(N)$ can be found in \citep{Reinhard07,Weissbluth12}.

$H_{DD}$ is responsible for the lifting of the degeneracy in the  $\{\ket{D,F,N-1},\ket{F,D,N-1}\}$ subspace \citep{Walker08} . The corresponding energies are represented by the parallel asymptotes of the blue lines in Fig. \ref{acoplados}. Notice the broad range of MW frequencies ($\sim 1.7$ GHz for the conditions of Fig. \ref{acoplados}) leading to eigenstates with the same total energy.\\

The coupling with the MW is described by:

\begin{eqnarray}
H_{MW}^{(2)} &=& \hbar \Omega_{MW}(\mathbf{\hat{z}_1}\otimes \mathds{1}_2+\mathds{1}_1\otimes \mathbf{\hat{z}_2})
\end{eqnarray}
where $\Omega_{MW}$ is the microwave Rabi frequency. Lower indeces refer to the corresponding atom.

$H_{MW}^{(2)}$ is responsible for the avoided crossings between the eigenenergies of $H_{DD}$.\\

Finally, $H_{vdW}$ represents the van der Waals energy shift of the $\ket{D_{\alpha},D_{\alpha^{\prime}}}$ states. Such shift results from second order coupling with states outside the basis set \citep{Reinhard07,Saffman10}. It is included phenomenologically as an isotropic interaction \citep{Amthor10}.

\begin{eqnarray}
H_{vdW} &=& \hbar\frac{C_{6}}{R^6}\sum_{\alpha,\alpha^{\prime}} \ket{D_{\alpha},D_{\alpha^{\prime}}}\bra{D_{\alpha},D_{\alpha^{\prime}}}
\end{eqnarray}

We have used $C_6 = 2\pi \times 15\ GHz.\mu m^6$ \citep{Reinhard07}.

\subsubsection{Optical excitation}

The interaction Hamiltonian for the optical excitation of an isolated Rydberg atom and a pair of Rydberg atoms are:

\begin{subequations}
\begin{align}
&V_{Opt}^{(1)} = \hbar \Omega_{Opt}  \mathds{1}_N \otimes\nonumber\\
&(\ket{{55}D_{\alpha_0}}\bra{g}e^{-i\omega^{\prime}t} +\ket{g}\bra{55D_{\alpha_0}}e^{i\omega^{\prime}t})\\
&V_{Opt}^{(2)} = \hbar \Omega_{Opt}  \mathds{1}_N \otimes\nonumber\\
&\left[(\ket{{55}D_{\alpha_0}}\bra{g}e^{-i\omega^{\prime}t} +\ket{g}\bra{55D_{\alpha_0}}e^{i\omega^{\prime}t})_1\otimes \mathds{1}_2\right.\nonumber\\
&+ \left.  \mathds{1}_1 \otimes e^{i\phi}(\ket{{55}D_{\alpha_0}}\bra{g}e^{-i\omega^{\prime}t}+ \ket{g}\bra{55D_{\alpha_0}}e^{i\omega^{\prime}t})_2 \right]\label{Vopt2}\nonumber\\
\end{align}
\end{subequations}
respectively. Here $()_i$ refers to the corresponding atom and $\mathds{1}_N$ is the identity operator in the MW Fock space. The factor $e^{i\phi}$ is included  to account for a possible optical field phase difference between positions 1 and 2 (it plays no role in transition probability calculations). 

\subsection{Numerical simulation}

The energies $e_i$ and eigenstates $\ket{v_i}$ of dressed single Rydberg atoms are calculated through the numerical diagonalization of the Hamiltonian $H^{(1)}$ restricted to the basis $\mathcal{B}^{(1)}(N) $.\\

Similarly, the energies $E_i$ and eigenstates $\ket{V_i}$ of a dressed Rydberg atom-pair, given the inter-atomic distance $R$ and the polar angle $\theta$, result from the diagonalization of the Hamiltonian $H^{(2)}$ restricted to the basis $\mathcal{B}^{(2)}(N) $. An example of the energy spectrum $\{E_i\}$ as a function of the MW frequency is shown in Fig. \ref{acoplados}.

In addition, the energies $E^{D}_i$ and eigenstates $\ket{V^{D}_i}$ are calculated through the diagonization of $H_0^{(2)}+H_{DD}$ (given $\{R,\ \theta\}$) to be employed in the calculation of the dipole - dipole potential energy of a given atom-pair state. \\ 

\begin{table*}
\begin{tabular}{|l|l|l|}
\hline
\textbf{Parameter}                           & \textbf{Value}                      & \textbf{Comment}                                                 \\ \hline
$\Gamma \quad ^{(*)}$               & $1.4 \times 10^4$ s$^{-1}$ & Typical value for Rydberg 55D state at room temperature \\ \hline
$\gamma \quad ^{(*)}$               & $14$ s$^{-1}$              & Order of magnitude of decay rate in Fig. \ref{transitorios}(ii)                                                        \\ \hline
$\gamma_0 \quad ^{(*)}$             & $1.4$ s$^{-1}$             & Order of magnitude of MOT charging/loss rate                                                       \\ \hline
$E_{th} \quad ^{(*)}$               & $h \times 0.14$ GHz        & Required DDI potential energy for atom expulsion in 100 $\mu$s (see Fig. \ref{expulsion})                                                        \\ \hline
$\Omega_{MW} $           & $2\pi \times 43$ MHz       &Estimated from amplifier-antenna system.                                                        \\ \hline
$\kappa$                            & $2\pi \times 11$ MHz       &Rydberg excitation decoherence rate. Estimated from laser performance.                                                         \\ \hline
$\Omega_{Opt}$                      & $2\pi \times 3.5$ MHz      &Adjusted parameter. Experimental estimate:  $2\pi \times 9.5$ MHz.                                                          \\ \hline
$\eta $                 & $68$ s$^{-1}$              &Adjusted parameter.                                                         \\ \hline
$a$                                 & $2.2\ \mu$m                &Adjusted parameter. In agreement with $a \sim \rho^{1/3}$.\\ \hline
$\bra{55D}\vert er \vert \ket{54F}$ & $5.7 \times 10^{-26}$ C m  & -6710 e a$_0$  \citep{Sibalic17}                                         \\ \hline
$C_6$                                & $15\ GHz.\mu m^6$          & \citep{Reinhard07,Beguin13}            \\ \hline
\end{tabular}
\caption{Values used in the numerical simulations shown in Figs. \ref{acoplados}, \ref{probabilidades} and \ref{espectroteorico}. The parameters signaled by $^{(*)}$ were given powers of ten values in units of the frequency difference between transitions 55D$_{3/2} \rightarrow$ 54F$_{5/2}$ and 56P$_{3/2} \rightarrow$ 55D$_{3/2} \simeq$ 1.4 GHz.}
\label{valoresparametros}
\end{table*}

\subsubsection{Optical excitation probabilities}

For an isolated atom initially in the ground state, the perturbed state $\ket{\psi}^{(1)}$ resulting from the coupling with the optical excitation was calculated using time-independent first order perturbation theory. It was used to determine
the transition amplitudes $a_i\equiv \braket{v_i |\psi}^{(1)}$  and the transition probabilities $p_i=\vert a_i \vert^{2}$. 
\begin{eqnarray}
a_i &=& \frac{\bra{v_i}V_{Opt}^{(1)} \ket{g}}{e_i-\Delta-i\kappa} \label{amplituddetransicionsimple}
\end{eqnarray}
where $\Delta$ is the optical detuning from the $5S_{1/2}(F=2)\rightarrow 55D_{3/2}$ transition. [$\Delta = 0$ in all simulations presented here] and $\kappa$ is an optical excitation dephasing rate.\\

\begin{figure}[ht!]
\centering
\includegraphics[width=1\linewidth]{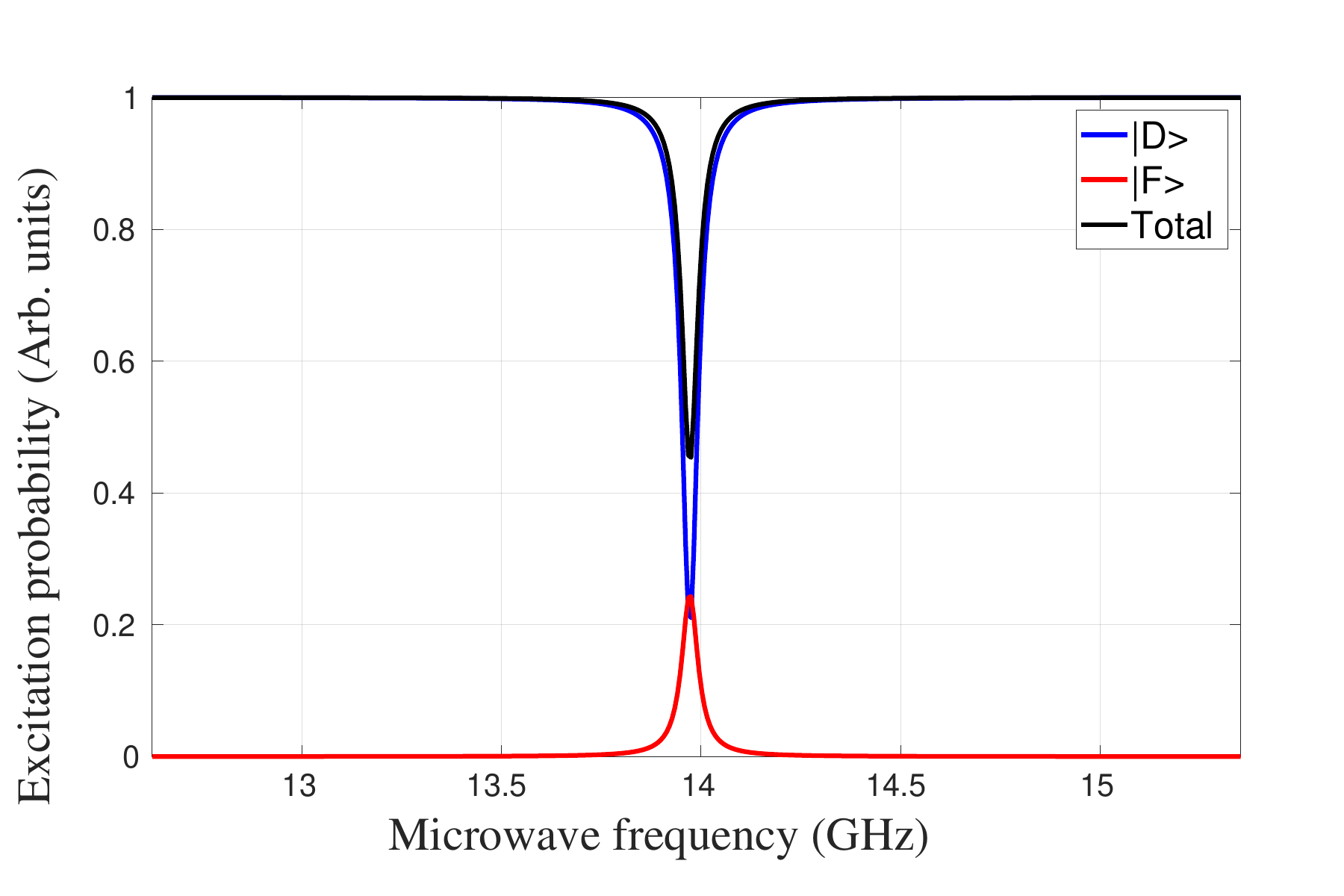}
\caption{\label{singles} Single atom Rydberg excitation probability as a function of the MW frequency ($\Omega_{MW} = 2\pi \times 43$ MHz, $\kappa = 2\pi \times 11$ MHz). }
\end{figure}

An example of the excitation probability for single atoms as a function of the MW frequency is shown in Fig. \ref{singles}. Notice the strong decrease of the excitation probability of  55D levels due to the avoided crossing of the dressed energies (AT effect) \cite{Tanasittikosol11,Brekke12}. On the other hand the excitation probability of level 54F is increased around the avoided crossing as a consequence of the mixing of $D$ and $F$ states.\\

In a similar way, the perturbed pair-state $\ket{\psi}^{(2)}$ was calculated using second order perturbation theory and used to compute the transition amplitudes  $A_j\equiv \braket{V_j|\psi}^{(2)}$ and the transition probabilities $P_j=\vert A_j \vert^{2}$.

\begin{eqnarray}
&A_j& = \sum_{i}\dfrac{\bra{V_j}V_{Opt}^{(2)} \ket{v_i,g}\bra{v_i,g}V_{Opt}^{(2)}\ket{g,g}}{(E_j-2\Delta-2i\kappa)(e_i-\Delta-i\kappa)}\nonumber\\
&+&\sum_{i}\dfrac{\bra{V_j}V_{Opt}^{(2)}\ket{g,v_i}\bra{g,v_i}V_{Opt}^{(2)}\ket{g,g}}{(E_j-2\Delta-2i\kappa)(e_i-\Delta-i\kappa)} \label{amplituddetransiciondoble}
\end{eqnarray}

A typical result for the double excitation probability for a given pair geometry is illustrated in see Fig. \ref{probabilidades}. A strong reduction at $\omega=\omega_{FD}$ occurs for the  Rydberg excitation probability of $\ket{D,D}$ pair states due to the AT effect mainly regarding the intermediate single-atom excitation. Also, the  excitation probability of $\ket{D,F}$ and $\ket{F,D}$  states is maximum at $\omega=\omega_{FD}$ and spreads over a broad MW frequency range corresponding to the many avoided crossings in Fig. \ref{acoplados}.

We signal that within the validity of perturbation theory the single atom excitation probability (fist order perturbation) is larger than double excitation probability (second order perturbation) by a factor $\mathcal{O}({\kappa}^2/\Omega_{Opt}^2)$.\\

\subsection{Transition rates}

The single excitation transition rate $\Lambda$ for transitions $gg \leftrightarrow gr$ was calculated from the single excitation transition amplitudes $a_i$ 
as:
\begin{eqnarray}
\Lambda= {\kappa}\sum_i \vert a_i \vert^{2}
\end{eqnarray}

Similarly, the double excitation rates were calculated from the transition amplitudes  $A_i\equiv \bra{V_i}V_{Opt}^{(2)}\ket{g,g}$ obtained through second order perturbation theory.
 
To discriminate between rates $\Lambda_1$ and $\Lambda_2$ we project each state $\ket{V_i}$ into the basis formed by the eigenstates $\ket{V^{D}_j}$ of  $H_0^{(2)}+H_{DD}$. Then the contribution of $\ket{V_i}$ to $\Lambda_1=\sum_{i}\Lambda_{1i}$ and $\Lambda_2=\sum_{i}\Lambda_{2i}$ are calculated as:
\begin{subequations}\label{expulsion}
\begin{eqnarray}
\Lambda_{1i} &=&{2\kappa} \vert A_i\vert^2 \sum_{E^{D}_{j}<E_{th}} \vert\braket{V^{D}_j\vert V_i}\vert^2\\
\Lambda_{2i} &=& {2\kappa}\vert A_i\vert^2 \sum_{E^{D}_{j}\geq E_{th}} \vert\braket{V^{D}_j\vert V_i}\vert^2
\end{eqnarray}
\end{subequations}
Where $E_{th}$ is the threshold potential energy considered sufficient to remove the atom-pair from the sample via the dipole - dipole force.

\subsection{Parameter values}

The parameter values used in the numerical simulations are given in Table \ref{valoresparametros}. Most parameters were given fixed values corresponding in order of magnitude to the experimental conditions. The parameters $\Omega_{Opt}$, $\eta$ and $a$ were varied for good qualitative agreement with the experimental spectra. A satisfactory agreement was obtained for values of $\Omega_{Opt}$ and $a$ close to the experimental estimates. 

\subsection{Kinematics of atom removal}

\begin{figure}[ht!]
\centering
\includegraphics[width=1\linewidth]{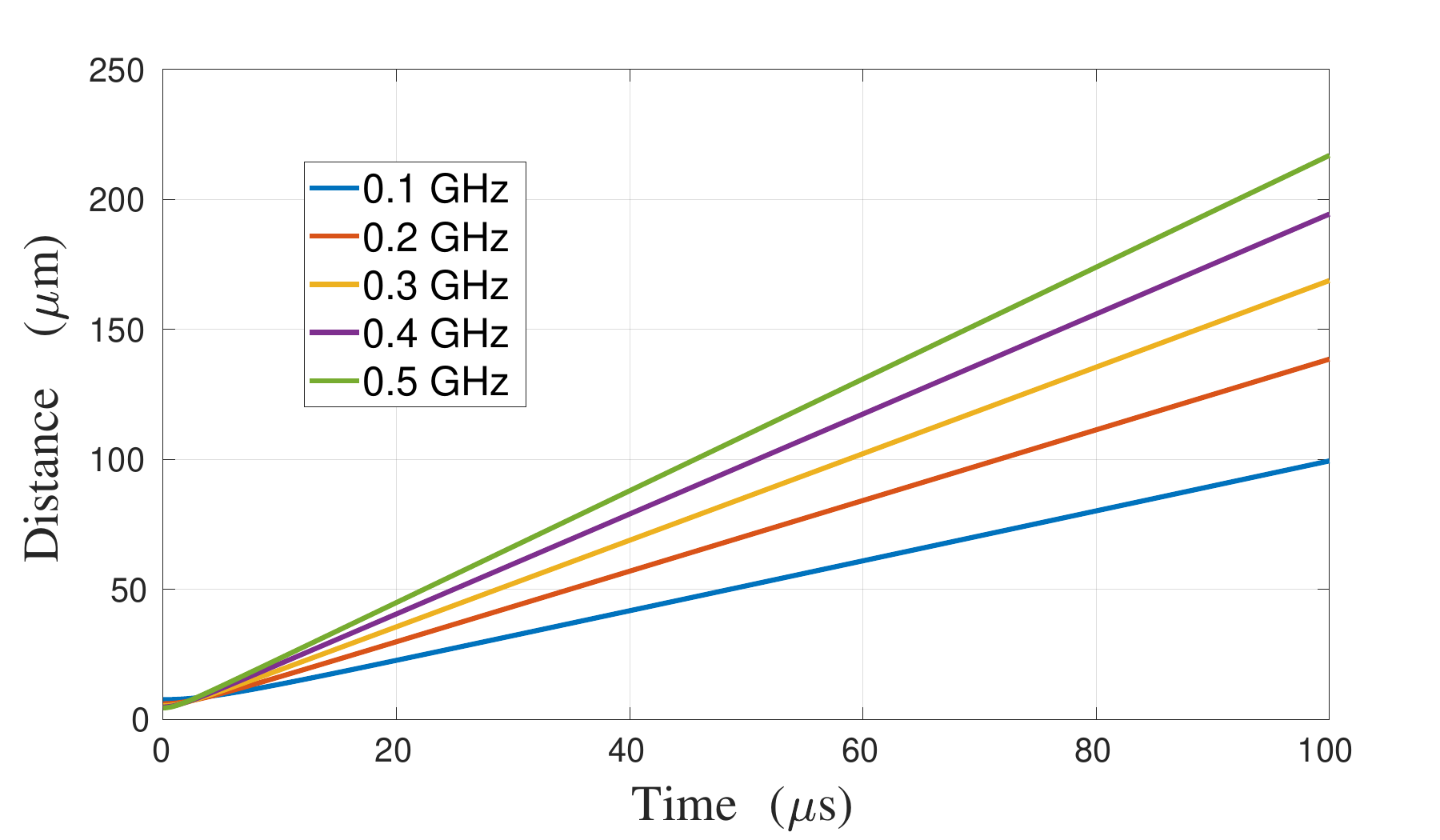}
\caption{\label{dxt} Evolution of the inter-atomic distance with time in response to the dipole - dipole force for several initial values of the potential energy. $C_3 = 43.7$ GHz.$\mu$m$^3$.}
\end{figure}

Figure \ref{dxt} shows the variation of the relative distance $R$ with time for an atom-pair submitted to a potential  $V(R)=C_3/R^3$ with $C_3= {\langle 55D\vert\vert er \vert\vert54F\rangle^2}/{4\pi\varepsilon_0 h}= 43.7$ GHz.$\mu$m$^3$ \citep{Sibalic17} for different values of the initial potential energy (in frequency units).

%\bibliographystyle{plain}
%\bibliographystyle{unsrt}
%\bibliography{Rydberg,RydbergH}

\end{document}